\documentclass{PoS}

\usepackage{amsmath}
\usepackage{amsbsy}
\title{Radiation from relativistic jets}

\ShortTitle{Radiation from relativistic jets}

\author{\speaker{K.-I. Nishikawa},$^a$ Y. Mizuno$^b$, 
P. Hardee$^c$, 
H. Sol$^d$, 
M. Medvedev$^e$, 
B. Zhang$^f$,
\AA. Nordlund$^g$,
J. T. Frederiksen$^g$,
G. J. Fishman$^h$, 
R. Preece$^a$,     \\                                     
\llap{$^a$} National Space Science and Technology Center,
320 Sparkman Drive, VP 62, Huntsville, AL 35805, USA\\
\llap{$^b$} National Space Science and Technology Center, 
320 Sparkman Drive, VP 62, Huntsville, AL 35805, USA, (visiting Univ. of Nevada, Las Vegas) \\
\llap{$^c$}  Department of Physics and Astronomy, The University of Alabama Tuscaloosa, AL 35487, USA \\
\llap{$^d$} LUTH, Observatore de Paris-Meudon, 5 place Jules Jansen, 92195 Meudon Cedex, France \\
\llap{$^e$} Department of Physics and Astronomy, University of Kansas, KS 66045, USA  \\
\llap{$^f$} Department of Physics, University of Nevada, Las
Vegas, NV 89154, USA  \\
\llap{$^g$}  Niels Bohr Institute, University of Copenhagen, Juliane Maries Vej 30,
2100 Copenhagen \O, Denmark  \\
\llap{$^h$} NASA-Marshall Space Flight Center, National Space Science and Technology Center
320 Sparkman Drive, VP 62, Huntsville, AL 35805, USA\\
E-mail: \email{ken-ichi.nishikawa-1@nasa.gov},
\email{yosuke.mizuno-1@nasa.gov}, \email{phardee@bama.ua.edu}, \email{Helene.Sol@obspm.fr}, \email{medvedev@ku.edu}, \email{bzhang@physics.unlv.edu}, 
\email{aake@astro.ku.dk}, \email{jacob@astro.ku.dk}, 
\email{jerry.fishman@nasa.gov}, \email{rob.preece@nasa.gov}}

\abstract{Nonthermal radiation observed from astrophysical systems containing relativistic jets and shocks, 
e.g., gamma-ray bursts (GRBs), active galactic nuclei (AGNs), and Galactic microquasar systems usually 
have power-law emission spectra. Recent PIC simulations of relativistic electron-ion (electron-positron) jets injected into a stationary medium show that particle acceleration occurs within the downstream jet. In the presence of relativistic jets, instabilities such as the Buneman instability, other two-streaming instability, and the Weibel (filamentation) instability create collisionless  shocks, which are responsible for particle (electron, 
positron, and ion) acceleration. The simulation results show that the Weibel instability is responsible for 
generating and amplifying highly nonuniform, small-scale magnetic fields. These magnetic fields contribute 
to the electron's transverse deflection behind the jet head. The ``jitter'' radiation from deflected electrons 
in small-scale magnetic fields has different properties than synchrotron radiation which is calculated in a 
uniform magnetic field. This jitter radiation, a case of diffusive synchrotron radiation, may be important to understand the complex time evolution 
and/or spectral structure in gamma-ray bursts, relativistic jets, and supernova remnants.
}

\FullConference{Workshop on Blazar Variability across the Electromagnetic Spectrum\\
		 April 22-25 2008\\
		 Palaiseau, France}

\begin{document}

\section{Introduction}
Shocks are believed to be responsible for prompt emission from gamma-ray 
bursts (GRBs) and their afterglows, for variable emission from blazars, 
and for particle acceleration processes in jets from active galactic
nuclei (AGN) and supernova remnants (SNRs). The predominant contribution
to the observed emission spectra is often assumed to be synchrotron- and 
inverse Compton radiation from these accelerated particles for gamma-ray bursts [1-7] and 
for AGN jets [8-13].
It is assumed that turbulent magnetic fields in the shock region 
lead to Fermi acceleration, producing higher energy particles \cite{fermi49,blaneich97}.
To make progress in understanding 
emission from these object classes, it is essential to place modeling efforts
on a firm physical basis. This requires studies of the microphysics of the 
shock process in a self-consistent manner \cite{p05b,wax06}. 

\section{Method of calculation}
Three-dimensional relativistic particle-in-cell (RPIC) simulations
have been used to study the microphysical processes in relativistic
shocks. Such PIC simulations show that rapid acceleration takes place
in situ in the downstream jet   
[18-33].
Three independent simulation
studies confirm that relativistic counter-streaming jets do excite
the Weibel instability \cite{weib59},   
which generates current filaments 
and associated magnetic fields \cite{ML99}, 
and accelerates electrons 
[18-22].

In order to determine the luminosity and spectral energy density (SED) 
of synchrotron radiation, it 
is general practise to simply assume that a certain fraction $\epsilon_{\rm B}$ 
of the post-shock thermal energy density is carried by the magnetic field, 
that a fraction $\epsilon_{\rm e}$ is carried by electrons, and that the 
energy distribution of the electrons is a power-law, $d \log n_{\rm e}/d
\log \varepsilon = p$ (above some minimum energy $E_{\rm m}$ which is
determined by $\epsilon_{\rm e}, \epsilon_{\rm B}$ and $p$). In this approach, 
$\epsilon_{\rm B}$, $\epsilon_{\rm e}$, and $p$ are treated as free parameters, 
to be determined by observations.  However, more constraining data now require additional 
free parameters such as the introduction of broken power-law to reproduce the spectral
energy distributions of TeV blazars for instance \cite{kata01}. 
Due to the lack of a first principle theory of collisionless shocks,  
a purely phenomenological approach to modeling radiation is 
applied, but one must recognize that emission is then calculated without
a full understanding of the processes responsible for particle acceleration 
and magnetic field generation \cite{wax06}. It is important to clarify that
the constraints imposed on these parameters by the observations are
independent of any assumptions regarding the nature of the
shocks and the processes responsible for particle acceleration or
magnetic field generation. Any model proposed for the actual shock
micro-physics must be consistent with these phenomenological constraints. 

Since magnetic fields are generated by the current structures produced
in the filamentation (Weibel) instability, 
it is possible that ``jitter'' radiation  [36-44] 
is an important emission process in
GRB and AGN jets. It should be noted that synchrotron- and  `jitter'-radiation are 
fundamentally the same physical processes (emission of accelerated
charges in a magnetic field), but the relative importance of the two  
regimes depends on the comparison of the deflection angle and the  
emission angle of the charges \cite{medv00}.
Emission via synchrotron- or ``jitter"-radiation from relativistic 
shocks is determined by the magnetic field strength and structure and 
the electron energy distribution behind the shock, which can be computed 
self-consistently with RPIC simulations. The full RPIC  
simulations may actually help to determine whether the emission is more  
synchrotron-like or jitter-like.

The characteristic differences between Synchrotron- and jitter radiation 
are relevant for a more fundamental understanding of the complex time evolution 
and/or spectral propertis of GRBs (prompt and afterglows) \cite{pre98}. 
For example, jitter 
radiation has been proposed as a solution of the puzzle that below their peak 
frequency GRB spectra are sometimes steeper than the ``line of death'' spectral 
index associated with synchrotron emission [35-38], 
i.e., the observed SED scales as $F_{\nu} \propto \nu^{2/3}$, whereas synchrotron 
SEDs should follow $F_{\nu} \propto \nu^{1/3}$, or even more shallow (i.e. $F_\nu \propto \nu^{\alpha}$ 
where $\alpha \leq 1/3$, e.g., \cite{medv06}). 
Thus, it is crucial to calculate the emerging radiation by tracing electrons 
(positrons) in self-consistently evolved electromagnetic fields. This highly 
complex analytical and computational task
requires sophisticated tools, such as multi-dimensional, relativistic, PIC methods.

\subsection{New Computing Method of Calculating Synchrotron
and Jitter Emission from Electron Trajectories in Self-consistently
Generated Magnetic Fields}

Consider a particle at position ${\bf{r}_{0}}(t)$ at time $t$ (Fig.\ 1). 
At the same time, we observe the associated electric field from position $\bf{r}$. 
Because of the finite propagation velocity of light, we actually observe the particle 
at an earlier position $\bf{r}_{0}(\rm{t}^{'})$ along its trajectory, labeled with
the retarded time $t^{'} = t - \delta t^{'} = t - \bf{R}(\rm{t}^{'})/c$. Here
$\bf{R}(\rm{t}^{'}) = |\bf{r} - \bf{r}_{0}(\rm{t}^{'})|$ is the distance from the 
charge (at the retarded time $t^{'}$) to the observer's position.

\begin{figure}[ht]
\begin{minipage}[t]{60mm}
\hspace*{0.8cm}
\includegraphics[width=.7\textwidth]{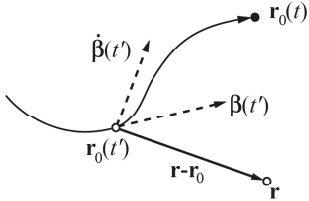}
\end{minipage}
\begin{minipage}[t]{77mm}
\vspace*{-2.5cm} \caption{\baselineskip 12pt Definition of the
retardation effect. From an observers point, r, one sees the particle at position 
$\bf{r}_{0}(\rm{t}^{'})$ where it was at retarded time t' (from Figure 2.2 in \cite{hedeT05}).} 
\end{minipage}
\end{figure}

The retarded electric field from a charged particle moving with
instantaneous velocity $\boldsymbol{\beta}$ under acceleration
$\boldsymbol{\dot{\beta}}$ is expressed as \cite{jack99},  

\vspace*{-0.7cm}
\begin{eqnarray}
\bf{E} & = & \frac{q}{4\pi\epsilon_{0}}
\left[\frac{\bf{n}-\boldsymbol{\beta}}{\gamma^{2}
(1-\bf{n}\cdot\boldsymbol{\beta})^{\rm{3}}\rm{R}^{2}}\right]_{\rm ret} +
\frac{q}{4\pi\epsilon_{0}c}
\left[\frac{\bf{n}\times\{(\bf{n}-\boldsymbol{\beta})\times
\boldsymbol{\dot{\beta}}\}}{(1-\bf{n}\cdot\boldsymbol{\beta})^{\rm{3}}\rm{R}}\right]_{\rm ret}
\end{eqnarray}

Here, $\bf{n} \equiv \bf{R}(\rm{t}^{'})/ |\bf{R}(\rm{t}^{'})|$ is a
unit vector that points from the particle's retarded position towards
the observer. The first term on the right hand side, containing the
velocity field, is the Coulomb field from a charge moving without
influence from external forces. The second term is a correction term
that arises when the charge is subject to acceleration. Since the
velocity-dependent field falls off in distance as $R^{-2}$, while the
acceleration-dependent field scales as $R^{-1}$, the latter becomes
dominant when observing the charge at large distances ($R \gg 1$).

The choice of unit vector $\bf{n}$ along the direction of propagation of
the jet (hereafter taken to be the $Z$-axis) corresponds to head-on emission. 
For any other choice of $\bf{n}$ (e.g., $\theta \lesssim 1/\gamma$), off-axis emission 
is seen by the observer. The observer's viewing angle is set by the choice of 
$\bf{n}$ ($n_{\rm x}^{2}+n_{\rm y}^{2}+n_{\rm z}^{2} = 1$). 
After some calculation and simplifying assumptions (for detailed
derivation see \cite{hedeT05}) 
the total energy $W$ radiated per unit
solid angle per unit frequency can be expressed as

\vspace*{-0.3cm}

\begin{eqnarray}
\frac{d^{2}W}{d\Omega d\omega} & = & \frac{\mu_{0} c
q^{2}}{16\pi^{3}} \left|\int^{\infty}_{-\infty}\frac{\bf{n}\times
[(\bf{n}-\boldsymbol{\beta})\times \boldsymbol{\dot{\beta}}]}{(1-\boldsymbol{\beta}\cdot
\bf{n})^{2}} e^{i\omega(t^{'} -\bf{n} \cdot \bf{r}_{0}({\rm t}^{'})/{\rm c})}
dt^{'}\right|^{2}
\end{eqnarray}

This equation contains the retarded electric field from a charged particle 
moving with instantaneous velocity $\boldsymbol{\beta}$ under acceleration
$\boldsymbol{\dot{\beta}}$, and only the acceleration field is kept since the
velocity field decreases rapidly as $1/R^{2}$. The distribution over frequencies
of the emitted radiation depends on the particle energy, radius of curvature, and 
acceleration. These quantities are readily obtained from the trajectory of each 
charged particle.

Since the jet plasma has a large velocity $Z$-component in the
simulation frame, the radiation from the particles (electrons and
positrons) is heavily beamed along the $Z$-axis as jitter radiation
\cite{medv00,medv06,flei06a,work08}.

\section{Radiation from relativistic electrons: a simple case to test computing method}
 
Here we have calculated the radiation from two electrons with Lorentz factor ($\gamma = 15.8, 40.8$)
\cite{nishi08a,nishi08b}.
 The electrons gyrate in the $x-z$ plane with the
uniform magnetic field ($B_{\rm y}$) and the results are shown in
Figures 2 \& 3.

\begin{figure*}[ht]
\centering
\vspace*{-0.7cm}
\includegraphics[width=0.24\linewidth]{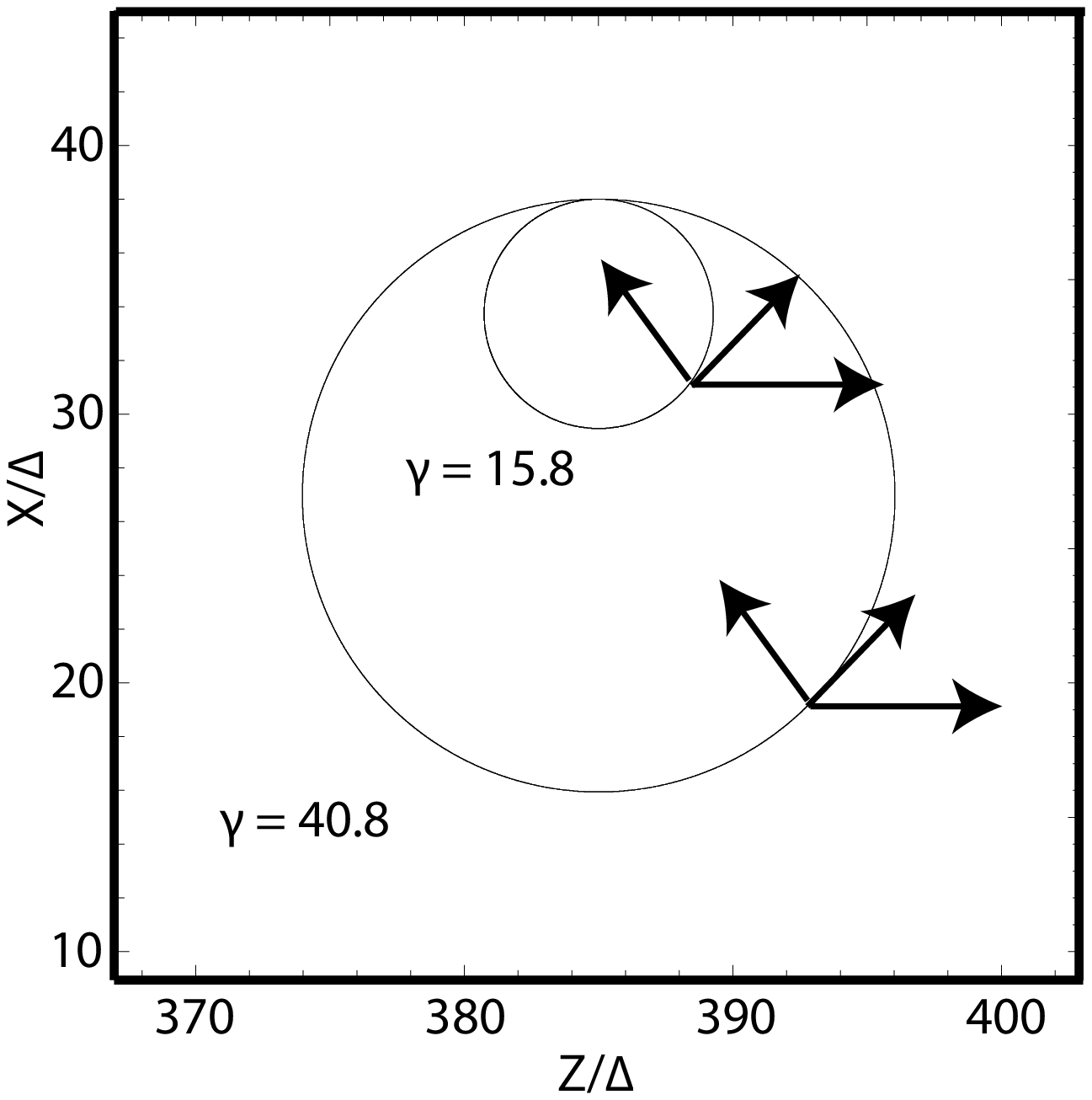}
\includegraphics[width=0.63\linewidth]{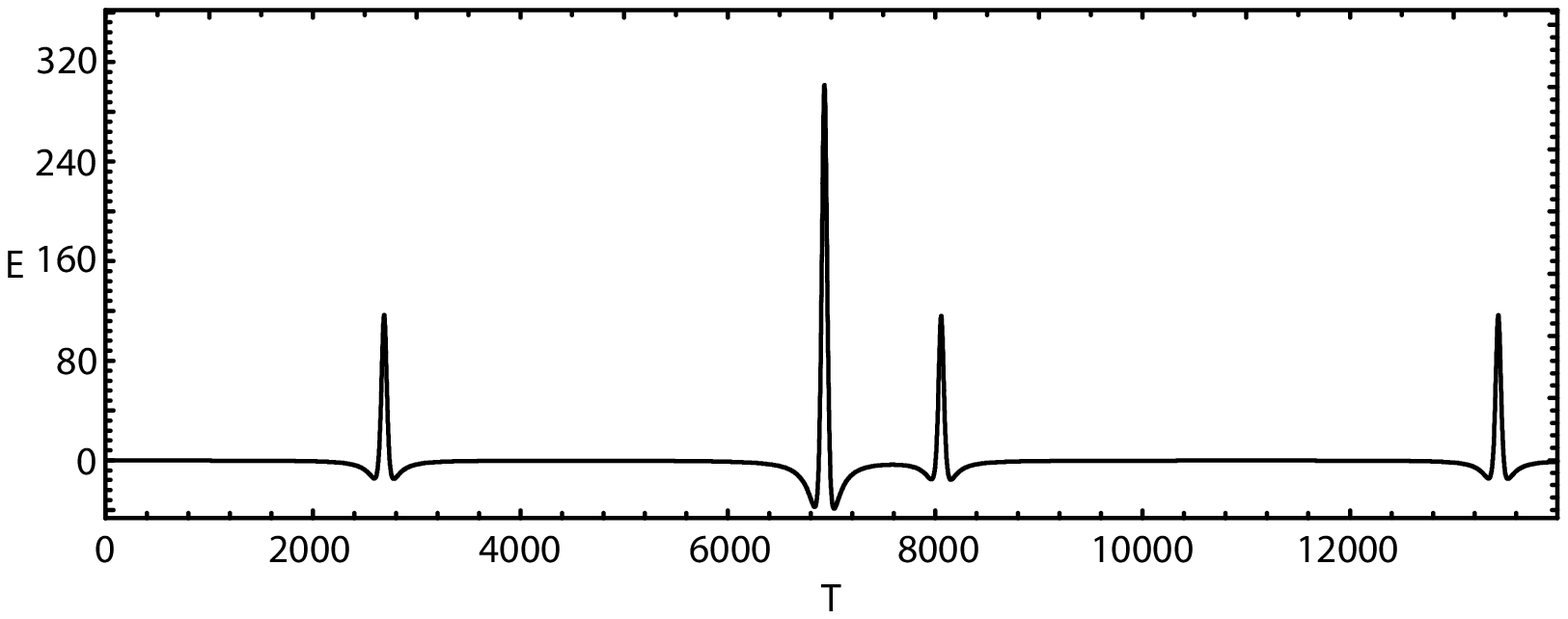}
\caption{\baselineskip 12pt The paths of two charged particles moving in a fixed homogenous
magnetic field (left panel) ($\gamma = 15.8, 40.8$). The particles produce a
time dependent electric field. An observer situated at great distance
along the n-vector sees the retarded electric field from the gyrating
particles (right panel). As a result of relativistic beaming, the field
is seen as pulses peaking when the particles move directly towards the
observer.}
\end{figure*}
\begin{figure*}[ht]
\centering
\begin{minipage}[t]{65mm}
\hspace*{-0.6cm}
\includegraphics[width=1.12\linewidth]{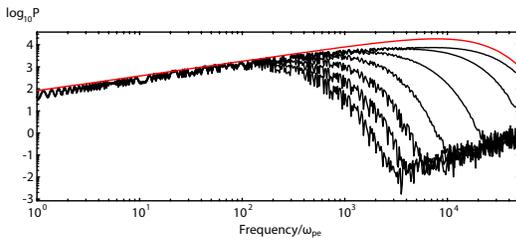}
\end{minipage}
\begin{minipage}[t]{78mm}
\hspace*{1.0cm}
\vspace*{-4.1cm}
\caption{\baselineskip 12pt The observed power spectrum from two charged particles,
gyrating in a magnetic field at different viewing angles. The
viewing angles are 0$^{\circ}$, 1$^{\circ}$, 2$^{\circ}$, 3$^{\circ}$,
4$^{\circ}$, 5$^{\circ}$, and 6$^{\circ}$ ($n_{\rm y}\ne 0$). With larger angles the frequencies above the Nyquist
frequency should be strongly damped, however they increase due to aliasing
\cite{hedeT05}. The units on both axes are arbitrary. 
The theoretical synchrotron spectrum for a viewing angle equal to 0$^{\circ}$
is plotted for comparison as a red curve for the electron with $\gamma = 40.8$
(multiplied by 2 for clarity).}  
\end{minipage}
\end{figure*}

The spectra observed far from the electron at angles
with respect to the $z$ direction are shown in Fig. 3. 
The higher frequencies ($> f_{\rm c}$) are strongly damped
with increasing angles as $e^{(-f/f_{\rm c})}$, see \cite{jack99}. 
Since the critical frequency $f_{\rm c} =
\frac{3}{2}\gamma^{3}\left(\frac{c}{\rho}\right) = 2309$, where $\rho
= 11.03$ for the electron with $\gamma = 40.8$ is larger than that for $\gamma = 15.8$,
the radiation from the electron $\gamma = 40.8$ is dominant \cite{nishi08b}. 
The electron with $\gamma = 15.8$ gyrates about three times in this period, the ripples in the spectrum shows
the electron cyclotron frequency. However, in order to resolve it much longer time is required \cite{hedeT05}.
We have very good agreement between the spectrum obtained from the simulation
and the theoretical synchrotron spectrum expectation (red curve) from eq.  3 
(eq. 7.10 \cite{hedeT05}).  

Synchrotron radiation with the full angular dependency for
the parallel polarization component is given by \cite{jack99}, 
\begin{eqnarray}
\frac{d^{2}W_{||}}{d\omega d\Omega} &=& \frac{\mu_{0}cq^{2}\omega^{2}}{12\pi}
\left( \frac{r_{\rm L}\theta^{2}_{\beta}\beta^{2}}{c}\right)^{2}\frac{\vert K_{\frac{2}{3}}
(\chi /\sqrt{\cos \theta \beta^{3}})\vert^{2}}{(\cos\theta\beta^{3})^{2}},
\end{eqnarray}

\noindent
where $\theta$ is the angle between {\bf n} and the orbital plane 
$\theta^{2}_{\beta}\equiv 2(1 - \beta\cos\theta)$, $\chi = \omega r_{\rm L}\theta^{3}_{\beta}/3c$ 
and $r_{\rm L}$ the gyro-radius $\gamma mv/(qB)$. For $\beta \rightarrow 1$ and $\theta \rightarrow 0$
this expression converges toward the solution one normally finds in text books \cite{jack99,rybic79}.

It should be noted 
that the method based on the integration of the retarded electric fields calculated
by tracing many electrons described in this section can provide a
proper spectrum in turbulent electromagnetic fields. On the other
hand, if the formula for the frequency spectrum of radiation emitted
by a relativistic charged particle in instantaneous circular motion
is used \cite{rybic79,jack99}, the complex particle
accelerations and trajectories are not properly accounted for and the
jitter radiation spectrum is not properly obtained (for details see 
\cite{hedeT05,hedeN05}). 
The results described above validate the technique used in our code as described previous section
\cite{nishi08a,nishi08b,hedeT05,hedeN05}.

\section{Discussion}

We have started to calculate emission directly from our
simulations using the same method described in the previous section. 
In order to calculate the (jitter-like) synchrotron
radiation from the particles in the electromagnetic fields generated
by the filamentation instability, the retarded electric field from a
single particle is Fourier-transformed to give the individual particle
spectrum as described in the previous section. The individual particle
spectra are added together to produce a total spectrum over a
particular simulation time span \cite{hedeT05,hedeN05}. 
It should be noted that for this calculation very large
simulations over a long time ($t_{\rm s}$) are required using a small
time step ($\Delta t$) in order to increase the upper frequency limit
to the spectrum (Nyquist frequency $\omega_{\rm N} = 1/2\Delta
t$). Frequency resolution is limited by the time span ($\Delta \omega
= 1/t_{\rm s}$) \cite{hedeT05,hedeN05}.   
For a case with the time step $\Delta
t = 0.01/\omega_{\rm pe}$ and the time span $t_{\rm s} =
50/\omega_{\rm pe}$, a calculated spectrum will have the highest
frequency, $50\omega_{\rm pe}$ and the frequency resolution (the
lowest frequency), $0.02\omega_{\rm pe}$.  $\omega_{\rm pe}$ is
calculated with an appropriate plasma density.  Simulations over a
long time allow us to obtain multiple spectra at sequential time spans
so the spectral evolution can be calculated. 
Synthetic spectra obtained
in the way we have described above should be compared with GRB prompt and afterglow observations.

In the case of AGN jets, diffusive synchrotron radiation has already 
been invoked by several works \cite{flei06b,FT07b,mao07} 
to reproduce spectra of 3C273, M87 
and Cen A knots from radio to X-rays. For TeV blazars, taking into account the relative 
importance of the energy densities contained in the small-scale and 
large-scale magnetic fields may be an elegant alternative to the choice 
of a broken power-law for the energy distribution of radiating electrons.

\acknowledgments
K.I.N., Y.M., and P.H. are  supported by NSF-AST-0506719, 
AST-0506666, NASA-NNG05GK73G, NNX07AJ88G, and NNX08AG83G. 
M.M has been supported by NSF-AST-0708213, NASA-NX07AJ50G, and NNX08AL39G, 
and DoE-DE-FG02-07ER54940.  Simulations were performed at the  
Columbia facility at the NASA
Advanced Supercomputing (NAS) and IBM p690 (Copper) at the National
Center for Supercomputing Applications (NCSA) which is supported by
the NSF. Part of this work was done while K.-I. N. was visiting the
Niels Bohr Institute. He thanks the director of the institution for 
generous hospitality. K.-I. N. prepared this proceeding during his visit at Meudon Observatory,
and  thanks the director of the institution for 
generous hospitality.

\end{document}